\documentclass{pasj00}

\SetRunningHead{M. Tsujimoto et al.}{Suzaku Observation of V458 Vul}

\Received{\today}
\Accepted{\today}

\begin{document}
\title{X-Ray Spectroscopy of the Classical Nova V458 Vulpeculae with Suzaku}
\author{
Masahiro~\textsc{Tsujimoto},\altaffilmark{1,2}\thanks{Chandra Fellow}
Dai~\textsc{Takei},\altaffilmark{3}
Jeremy~J.~\textsc{Drake},\altaffilmark{4}
Jan-Uwe~\textsc{Ness},\altaffilmark{5}\thanks{Chandra Fellow}
and Shunji~\textsc{Kitamoto}\altaffilmark{3}
}
\altaffiltext{1}{Department of Astronomy \& Astrophysics, Pennsylvania State University,\\
525 Davey Laboratory, University Park, PA 16802, USA}
\altaffiltext{2}{Japan Aerospace Exploration Agency, Institute of Space and
Astronautical Science,\\
3-1-1 Yoshino-dai, Sagamihara, Kanagawa 229-8510}
\altaffiltext{3}{Department of Physics, Rikkyo University, 3-34-1 Nishi-Ikebukuro, Toshima, Tokyo 171-8501}
\altaffiltext{4}{Smithsonian Astrophysical Observatory, MS-3, 60 Garden Street, Cambridge, MA 02138, USA}
\altaffiltext{5}{School of Earth and Space Exploration, Arizona State University, Tempe, AZ 85287, USA}
\email{tsujimot@astro.isas.jaxa.jp}
\maketitle

\begin{abstract}
 We conducted a target of opportunity X-ray observation of the classical nova V458
 Vulpeculae 88 days after the explosion using the Suzaku satellite. With a $\sim$20~ks
 exposure, the X-ray Imaging Spectrometer detected X-ray emission significantly harder
 than typical super-soft source emission. The X-ray spectrum shows K$\alpha$ lines from
 N, Ne, Mg, Si, and S, and L-series emission from Fe in highly ionized states. The
 spectrum can be described by a single temperature ($\sim$0.64~keV) thin thermal plasma model
 in collisional equilibrium with a hydrogen-equivalent extinction column density of
 $\sim 3 \times 10^{21}$~cm$^{-2}$, a flux of $\sim 10^{-12}$~erg~s$^{-1}$~cm$^{-2}$,
 and a luminosity of $\sim$6$\times$10$^{34}$~erg~s$^{-1}$ in the 0.3--3.0~keV band at
 an assumed distance of 13~kpc. We found a hint of an enhancement of N and deficiencies
 of O and Fe relative to other metals. The observed X-ray properties can be interpreted
 as the emission arising from shocks of ejecta from an ONe-type nova.
\end{abstract}

\section{Introduction}\label{s1}
Classical novae occur in accreting binaries with a white dwarf (WD) as the
primary. Hydrogen-rich accreted material is accumulated on the WD surface until a
critical mass is reached that ignites a thermonuclear runaway. A wind driven by
radiation pressure leads to ejection of accreted and partially nuclear-processed
material, forming an optically thick shell around the binary system. The nova terminates
once the remaining hydrogen is consumed on the WD surface. Reviews of classical novae
can be found in e.g., \citet{gehrz98,starrfield08}.

Classical novae are copious X-ray emitters (e.g., \cite{krautter02}). Several processes
are involved in producing the X-rays. One is photospheric emission from a hot layer of
the WD surface fueled by residual nuclear burning after the explosion. The layer is
rendered visible after the ejecta shell expands and becomes less opaque to soft
X-rays. Its X-ray spectrum is characterized by a very soft blackbody-like emission with
a temperature of $\stackrel{<}{_\sim}$50~eV and is easily identified in
medium-resolution CCD spectra. Higher-resolution grating spectroscopy can further
identify absorption features by the WD atmosphere over the blackbody continuum
\citep{ness03}. These types of X-ray emitters are known as super-soft sources (SSS).

Another type of X-ray emission originates from circumstellar material that is
photoionized by the SSS emission. This reprocessed emission is usually dwarfed by the
ionizing source. However, when the surface nuclear fuel is consumed and the SSS emission
fades out, residual emission from the heated surrounding medium can emerge. This was
found in the nova V382 Vel, in which the residual emission was interpreted as the
radiatively cooling ejecta that are optically thin and have reached a collisional
equilibrium \citep{ness05}. Similar line-dominated emission was found in V4743 Sgr, when
the SSS emission turned off for a brief period for unknown reasons \citep{ness03}.

Yet another type is weak and hard X-rays, which a substantial fraction of X-ray-emitting
novae exhibit within a year of the outburst before the SSS emission becomes visible. In
systematic studies by ROSAT \citep{orio01} and Swift \citep{ness07}, more than half and
six out of eight are among such novae, respectively. While their spectra are similar to
that found in V382 Vel \citep{ness05}, their origin is probably different; the ejected
material has not expanded enough to be optically thin, and no ionizing source is present
to heat up the surrounding material. Several possible origins have been proposed for the
hard X-rays including shock heating of expanding ejecta \citep{lloyd92,obrien94} and
reestablished accretion. The latter explains the X-rays seen in a nova by a magnetic WD
(V2487 Oph; \cite{hernanz02}).

Detailed and timely spectroscopy is requisite to understanding the origin of the early
hard X-ray emission, but it is complicated by the faint and more transient nature of
this phenomenon. While grating spectroscopy yields a wealth of information, it is
limited to extremely bright sources. Only two novae have been studied with grating
spectroscopy for their hard emission, but they are atypical samples; RS Oph
\citep{nelson08,drake08a,ness08a} is a symbiotic nova, where shocks are expected between
the expanding ejecta and the stellar wind of the giant companion. Hard emission from
V382 Vel \citep{ness05} was also observed with grating, but the emission has a different
origin as described earlier. Hard emission in most classical novae with a main sequence
companion has generally been studied based on scant low-resolution spectra and data of
poor statistical quality. The utility of such data is limited to determination of the
plasma temperature and X-ray extinction from broad-band spectral shapes. Few studies to
date attempted to go beyond this \citep{mukai01,orio01,hernanz07}.

Suzaku fills the gap well. Its medium-resolution CCD spectrometer yields a high
signal-to-noise ratio spectrum for moderately bright X-ray sources in a reasonable
telescope time. It has sufficient spectral resolution to resolve emission lines from
abundant elements. While we can only rely on global spectral models, the relative flux
and energy of these lines enable us to conduct a temperature diagnosis, to study the
plasma evolution using non-equilibrium features in emission lines, and to reveal
abundance patterns in the X-ray-emitting plasma.

Here, we present a Suzaku spectroscopic study of the nova V458 Vulpeculae (V458 Vul) 88
days after the explosion. A high signal-to-noise ratio spectrum of the hard X-ray
emission was obtained using the X-ray Imaging Spectrometer (XIS). Emission lines from
highly-ionized N, Ne, Mg, Si, S, and Fe are resolved, enabling us to conduct the types
of diagnosis described above. We present the finding of the study and discuss their
implications.

\section{V458 Vulpeculae}\label{s2}
\subsection{Ground-based monitoring}
\begin{figure}
 \begin{center}
  \FigureFile(85mm,85mm){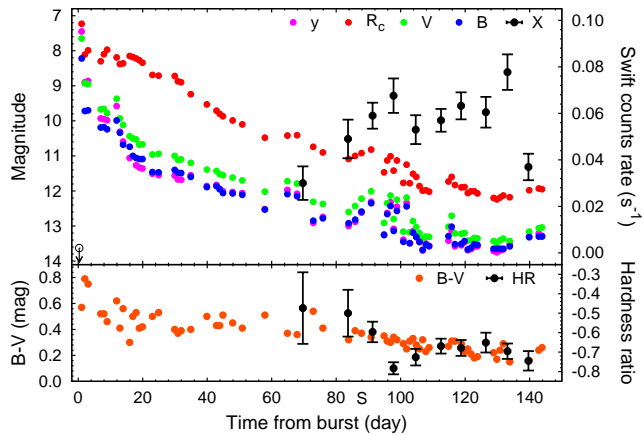}
 \end{center}
 \caption{Evolution of V458 Vul in brightness and color (upper and lower panels,
 respectively). The origin of the abscissa is day 54320.54 in modified Julian date when
 the nova was first spotted \citep{nakano07}. The epoch of the Suzaku observation is
 indicated with ``S'' on the abscissa. In the upper panel, the \textit{y}-,
 \textit{R\rm{c}}-, \textit{V}-, and \textit{B}-band magnitudes (private communication
 with K. Nakajima) are respectively shown in magenta, red, green, and blue, while the
 Swift X-ray count rate (0.3--10.0~keV) is shown in black with 1 $\sigma$ errors
 \citep{ness08b}. The 95.5\% upper-limit for non-detected X-rays on day 1 is indicated
 by an open circle. In the lower panel, the \textit{B}--\textit{V} color is shown in
 orange, while the Swift hardness ratio is in black with 1 $\sigma$ errors. Hardness
 ratio is defined by (H$-$S)/(H$+$S), where S and H are the Swift count rates in the
 0.25--1.5~keV and 1.5--10~keV band, respectively.}\label{f1}
\end{figure}

On 2007 August 8.54 UT, an optical nova reaching a visual magnitude of $\sim$9.5~mag was
spotted in the constellation Vulpecula at (RA, Dec) $=$ (\timeform{19h54m24.66s},
\timeform{+20D52'51.7''}) in the equinox J2000.0 \citep{nakano07}. The variable star was
fainter by $\sim$8~mag before the explosion if a cataloged source (USNO-A2.0
1050-15545600; \cite{monet98}) is its precursor. Optical spectroscopic observations on
the following day revealed hydrogen and helium emission lines with P-Cygni profiles with
a full width at half maximum (FWHM) of 1750--1900~km~s$^{-1}$ \citep{buil07},
establishing the classical nova nature of this source. The nova was named V458 Vul
\citep{samus07}.

Subsequent ground-based observations were made both photometrically and
spectroscopically. The light curve of the nova
\citep{bianciardi07a,bianciardi07b,broens07,casas07,labordena07,nakamura07,scarmato07}
declined from the maximum \textit{V}-band magnitude of 7.65~mag on day 1.42 to
$\sim$14~mag in five months with time scales of $t_{2}\sim$~7~day \citep{poggiani08},
where $t_{2}$ is the time of decline from the optical maximum by 2 mag. The declining
trend is globally monotonic, but with several exceptions of local brightening at around
68 and 95 days after the explosion (figure~\ref{f1}). The earliest spectra
\citep{kiss07,lynch07,munari07,prater07,skoda07,tarasova07} are characterized by P Cygni
profiles in H Balmer series, He\emissiontype{I}, and Fe\emissiontype{II} lines. From all
these data, the nova was initially classified as a very fast \citep{payne57} and a
Fe\emissiontype{II}-type \citep{williams92} nova. However, subsequent observations
obtained approximately one month after the outburst show that the Fe lines have faded
and the N\emissiontype{II} and N\emissiontype{III} lines have gained prominence,
indicating an evolution toward a He/N class nova \citep{poggiani08}. \cite{wesson08}
found that this nova occurred inside a planetary nebula with H$\alpha$ imaging, a second
such case after GK Per.

We adopt the distance to V458 Vul to be 13~kpc \citep{wesson08}. In a series of
H$\alpha$ images taken in May 2008, a bright circumstellar knot was discovered at
3\farcs5\ away from the central source. By assuming that the knot is illuminated by the
flash of the nova, the distance of $\sim$13~kpc was derived. This is consistent with two
other distance estimates. The one is derived from a maximum-magnitude versus
rate-of-decline relation \citep{downes07}, an empirical relation that intrinsically
bright novae fade our fast. The other is derived from the assumption that the measured
radial velocity stems only from the Galactic rotation \citep{wesson08}.

The interstellar extinction ($A_{V,\rm{ISM}}$) was estimated to be 1.76~$\pm$~0.32~mag
\citep{poggiani08}. This converts to an interstellar hydrogen-equivalent extinction
column density ($N_{\rm H,ISM}$) of $3.15 \pm 0.57 \times 10^{21}$~cm$^{-2}$ using the
relation $N_{\rm{H,ISM}}$/$A_{V,\rm{ISM}}= 1.79 \times 10^{21}$~cm$^{-2}$~mag$^{-1}$
\citep{predehl95}. The line-of-sight extinction integrated through our Galaxy toward
V458 Vul is estimated from H\emissiontype{I} maps to be 3.7$\times$10$^{21}$~cm$^{-2}$
\citep{kalberla05} or 4.0$\times$10$^{21}$~cm$^{-2}$ \citep{dickey90}, which is
consistent that the nova is within in our Galaxy.

\subsection{Space-based monitoring}
Monitoring observations were also conducted with space-based facilities. The X-Ray
Telescope (XRT; \cite{burrows05}) aboard Swift was employed to take an immediate
follow-up image on day 1 and repeated snapshots from day 71 to 140
(figure~\ref{f1}). The X-ray emission was not detected on day 1, but it started to
emerge on day 70 during the first optical rebrightening. A total of 192 X-ray counts
were accumulated in a short observation of $\sim$6.7~ks, which is at least 10 times
brighter than the upper limit to the X-ray counts on day 1 \citep{drake07}. The X-ray
flux continued to increase until the peak on day $\sim$100, showed a stable flux for
$\sim$30 days, then decreased in flux quite sharply on the last Swift visit on day 140.

The XRT spectrum on day 70 has significant signal above 1~keV, indicating that hard
X-ray emission component was present in addition to any soft emission typical of the SSS
phase \citep{drake07}. No detailed spectroscopy was possible due to the paucity of
counts. We therefore requested a director's discretionary time on the Suzaku telescope
to obtain an X-ray spectrum of better statistics and resolution. A $\sim$20~ks
observation was conducted on day 88. We present the detailed analysis of the Suzaku
observation in this paper. Swift resumed its observing campaign on day 315. In this
later phase, the hard X-ray flux is highly variable and is anti-correlated with the UV
flux. The Swift results, including the later observations, are presented in
\citet{drake08b,ness08b}.

\section{Observation}\label{s3}
Suzaku observed V458 Vul on 2007 November 4 from 7:51 UT to 21:00 UT. The observatory
\citep{mitsuda07} carries two instruments: the X-ray Imaging Spectrometer (XIS:
\cite{koyama07}) sensitive at 0.2--12~keV and the Hard X-ray Detector (HXD:
\cite{kokubun07,takahashi07}) at 10--600~keV. We concentrate on the XIS data in this
paper, which has sensitivity for the energy range of interest.

The XIS is equipped with four X-ray CCDs. Three of them (XIS0, 2, and 3) are
front-illuminated (FI) devices, while the remaining one (XIS1) is back-illuminated (BI).
FI and BI CCDs are superior to each other in the hard and soft band responses,
respectively. XIS2 has not been functioning since 2006 November, and we use data from
the remaining three devices. The CCDs are mounted at the focus of four independent X-ray
telescopes \citep{serlemitsos07}, which are aligned to observe a $\sim$18\arcmin
$\times$18\arcmin\ field with a half-power diameter of $\sim$2\arcmin.

The three imaging-spectrometers have a combined effective area of $\sim$1030~cm$^{2}$ at
1.5~keV. We used the spaced-row charge injection technique, which rejuvenates the
degraded spectral resolution by filling the charge traps with artificially injected
electrons during CCD readouts. This yields an FWHM resolution of 145--174~eV at 5.9~keV.

The observation was conducted using the normal clocking mode with a frame time of
8~s. Data were cleaned using processing version 2.1 to remove non--X-ray events and
those taken during the South Atlantic Anomaly passages, at elevation angles below
5$^\circ$ from the Earth rim, and at elevation angles below 20$^\circ$ from sunlit Earth
rim. After filtering, the net integration time is $\sim$19.6~ks. We used HEASoft version
6.4.1\footnote{See http://heasarc.gsfc.nasa.gov/docs/software/lheasoft/ for detail.} for
the data reduction and Xspec version 11.3.2\footnote{See
http://heasarc.gsfc.nasa.gov/docs/xanadu/xspec/index.html for detail.} for the X-ray
spectral analysis.

\section{Analysis}\label{s4}
\subsection{Image and Timing}
\begin{figure}
 \begin{center}
  \FigureFile(85mm,85mm){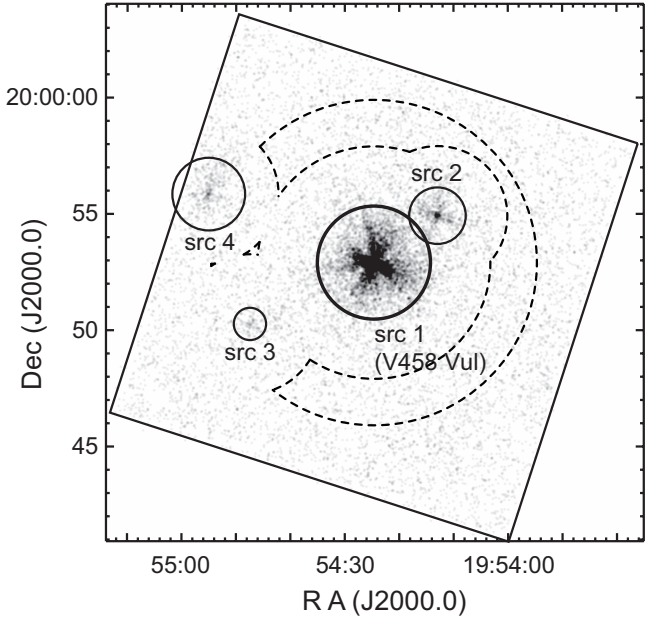}
 \end{center}
 \caption{Smoothed XIS image in the 0.2--5.0~keV band. Three CCD images were merged and
 a Gaussian smoothing was applied. V458 Vul (source 1) and three faint sources (sources
 2--4) were detected. The source regions are shown with solid circles around each
 source, while the background region for V458 Vul is shown with broken annulus with
 masks.}\label{f2}
\end{figure}

Figure~\ref{f2} shows the XIS image in the 0.2--5.0~keV band obtained by merging the
events recorded by all the three CCDs. Four sources were detected by visual
inspection. The brightest one at the center is V458 Vul based on astrometric consistency
with the Swift data. We fine-tuned the coordinate of the XIS image using this source as
a reference. The resulting positions, X-ray count rates (CR), and hardness ratio (HR) of
the four sources are summarized in table~\ref{t1}. None of the sources except for V458
Vul has records in the SIMBAD or the NASA/IPAC Extragalactic Database, leaving their
natures unclear.

\begin{table}
 \begin{center}
  \caption{Source list.}\label{t1}
  \begin{tabular}{lcccc}
   \hline 
   Source & RA & Dec & CR\footnotemark[$*$] & HR\footnotemark[$\dagger$] \\
   & (J2000.0) & (J2000.0) & (s$^{-1}$) & \\
   \hline
   1\footnotemark[$\ddagger$] & \timeform{19h54m25s} & \timeform{+20D52'52''} & 9.8$\times$10$^{-2}$ & $-$0.55 \\
   2                          & \timeform{19h54m13s} & \timeform{+20D54'53''} & 7.9$\times$10$^{-3}$ & $+$0.43 \\
   3                          & \timeform{19h54m47s} & \timeform{+20D50'13''} & 2.7$\times$10$^{-3}$ & $+$0.40 \\
   4                          & \timeform{19h54m55s} & \timeform{+20D55'48''} & 5.2$\times$10$^{-3}$ & $-$0.34 \\
   \hline
   \multicolumn{4}{@{}l@{}}{\hbox to 0pt{\parbox{85mm}{\footnotesize
   \par\noindent
   \footnotemark[$*$] The background-subtracted count rate in the 0.2--5.0~keV
   averaged over the three CCDs. The values are normalized to the circular aperture of
   a 3\arcmin\ radius.
   \par\noindent
   \footnotemark[$\dagger$] The hardness ratio defined as (H$-$S)/(H$+$S), where H and
   S are background-subtracted count rates in the hard (1.5--5.0~keV) and soft
   (0.2--1.5~keV) band, respectively.
   \par\noindent
   \footnotemark[$\ddagger$] V458 Vul. The contamination by source 2 is not corrected in the
   CR and HR values.
  }\hss}}
  \end{tabular}
 \end{center}
\end{table}

The background was subtracted to compute CR and HR. Source counts were integrated in
circles with adaptively-chosen radii to maximize the ratio against background for all
sources (solid circles in figure~\ref{f2}). Background counts were accumulated in a
different method for V458 Vul and the others. For V458 Vul, background was from an
annular region with inner and outer radii of 5\arcmin\ and 7\arcmin, respectively. For
the other sources, it was from annuli around V458 Vul with the inner circle
circumscribed to and the outer circle inscribed to the source extraction circle of each
source. The 3\arcmin\ circles around all sources were masked in these annuli. In this
way, we canceled the contamination to the faint three sources by the dominantly bright
source 1. The background was corrected for the vignetting before subtracting from the
source.  The contamination of source 1 by source 2 is considered later.

We constructed a light curve of V458 Vul and found no significant changes. The curve was
fitted with a constant flux model with a null hypothesis probability of 53\%.

\subsection{Spectrum}\label{s4-2}
\begin{figure}
 \begin{center}
  \FigureFile(85mm,85mm){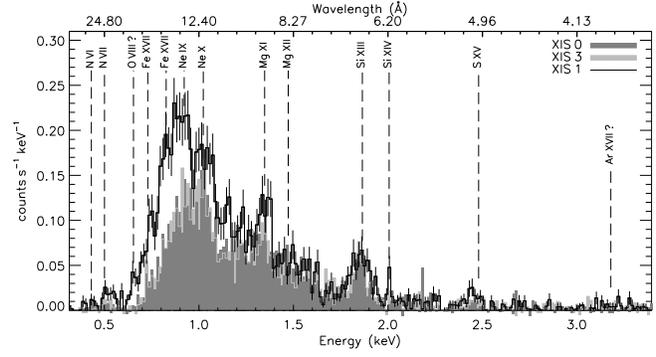}
 \end{center}
 \caption{X-ray spectra of V458 Vul (background subtracted) in a linear scale observed
 with the XIS0 and XIS3 (both FI devices; dark and light gray shadings, respectively)
 and with the XIS1 (BI device; solid line). The energies of plausible emission lines are
 labeled.}\label{f3}
\end{figure}

Figure~\ref{f3} shows the background-subtracted XIS spectra of V458 Vul in a linear
scale at the 0.4--3.5~keV band. We see conspicuous K$\alpha$
line emission from N\emissiontype{VII}, Ne\emissiontype{IX}, Ne\emissiontype{X},
Mg\emissiontype{XI}, Mg\emissiontype{XII}, Si\emissiontype{XIII}, and
S\emissiontype{XV}. We also see some emission from Fe\emissiontype{XVII} L-series lines.

\begin{figure}[ht!]
 \begin{center}
  \FigureFile(85mm,85mm){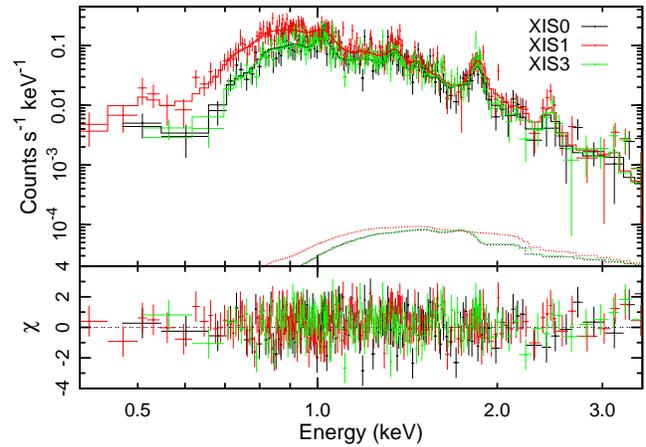}
 \end{center}
 \caption{Best-fit models to the background-subtracted spectra in a logarithmic
 scale. Different colors are used for each CCD. The top panel shows the data with
 crosses and the best-fit model with solid lines. The contamination by the nearby source
 2 is included as dotted lines. The bottom panel shows the residuals to the
 fit.}\label{f4}
\end{figure}

In order to model the spectrum, we generated the redistribution matrix functions and the
auxiliary response functions using \texttt{xisrmfgen} and \texttt{xissimarfgen}
\citep{ishisaki07}, respectively. The difference in effective area arising from
different off-axis angles between the source and the background regions were compensated
following the procedure described in \citet{hyodo08}.

The contamination by the nearby source 2 was considered by including an additional
absorbed power-law model (7.8$\times$10$^{21}$~cm$^{-2}$ for the absorption column and
2.1 for the photon index) that best describes its spectrum. From a ray-tracing
simulation, we found that $\sim$1.5\% of the emission from source 2 is in the source
extraction region of source 1. Fortunately, sources 1 and 2 have different spectral
hardness (table~\ref{t1}) and the contamination by source 2 is almost negligible.

We fitted the 0.3--3.0~keV spectrum of source 1 using an optically-thin thermal plasma
model in collisional equilibrium (the APEC model; \cite{smith01}) convolved with an
interstellar absorption assuming the cross sections and the chemical abundance in
\citet{wilms00}. The plasma temperature ($k_{\rm{B}}T$) and its emission measure as well
as the hydrogen column density ($N_{\rm H}$) were allowed to vary.

For the chemical composition of the plasma, we tried four models with different constraints
on the abundance of elements (Model 1--4). The base-line model is the ``Model 1'', in
which the abundance of elements with noticeable K$\alpha$ emission (N, Ne, Mg, Si, and
S) were allowed to vary individually. Also, the abundance of Fe was also varied, as the
Fe L series emission is necessary to explain the spectrum between the
O\emissiontype{VIII} K$\alpha$ (0.65~keV) and Ne\emissiontype{IX} K$\alpha$ (0.92~keV)
features. The abundance of Ni, which also contributes through its L series emission, was
tied to that of Fe at the solar ratio. Other elements were fixed to the solar
abundance. Here, we assumed the solar values by \citet{anders89}. The spectrum and the
best-fit model is shown in figure~\ref{f4}, and the best-fit parameters are summarized
in table~\ref{t2}. Additional components are not statistically required to explain the
spectra.

For the other models, we put additional constraints on the ``Model 1'' for the purpose
of reducing the statistical uncertainties. In the ``Model 2'', we tied the abundance of
intermediate-mass elements (Ne, Mg, Si, S) after confirming that the abundance of these
elements is consistent with each other in the ``Model 1`'. In the ``Model 3'', we fixed
the N abundance to be 1 solar. In the ``Model 4'', we fixed the abundance of less
constrained elements (N, O, and S) than the others to be 1 solar. The best-fit
parameters for these models are also tabulated in table~\ref{t2}. The results are
consistent among all these models except for the ''Model 4``, in which the estimate of
$N_{\rm{H}}$ and the abundance values are higher and the $\chi^{2}$-fitting is worse
than the others with significant residuals in the N--O K$\alpha$ energy range. We
therefore consider that this is not an appropriate model, and interpret the spectrum
based on the result of the other three models.

\begin{table*}[th!]
 \begin{center}
  \caption{Best-fits spectral parameters.}\label{t2}
  \begin{tabular}{llcccc}
   \hline
   Param. & Unit & Model 1\footnotemark[$*\dagger$] & Model 2\footnotemark[$*\dagger$] & Model 3\footnotemark[$*\dagger$] & Model 4\footnotemark[$*\dagger$] \\
   \hline
   $N_{\rm{H}}$  & 10$^{21}$ (cm$^{-2}$) \dotfill 
       & 2.8 (1.8--4.6) & 2.8 (1.8--4.4) & 2.8 (1.9--4.1) & 3.8 (2.9--4.8) \\
   $k_{\rm{B}}T$ & (keV)   \dotfill 
       & 0.64 (0.57--0.71) & 0.65 (0.60--0.71) & 0.64 (0.60--0.70) & 0.64 (0.59--0.69) \\
   $Z_{\rm{N}}$  & (solar) \dotfill 
       & 1.0 (0.0-45) & 1.3 (0.0--35) & 1 (fixed) & 1 (fixed) \\
   $Z_{\rm{O}}$  & (solar) \dotfill 
       & 0.0 (0.0--1.6) & 0.0 (0.0--1.1) & 0.0 (0.0--0.8) & 1 (fixed) \\
   $Z_{\rm{Ne}}$ & (solar) \dotfill 
       & 0.6 (0.3--1.8) & 0.5 (0.3--1.1) & 0.6 (0.3--1.2) & 1.2 (0.6--1.9) \\
   $Z_{\rm{Mg}}$ & (solar) \dotfill 
       & 0.4 (0.2--1.3) & 0.5 (0.3--1.1) & 0.4 (0.2--0.9) & 0.9 (0.5--1.4) \\
   $Z_{\rm{Si}}$ & (solar) \dotfill 
       & 0.5 (0.3--1.3) & 0.5 (0.3--1.1) & 0.5 (0.3--0.9) & 0.9 (0.5--1.5) \\
   $Z_{\rm{S}}$  & (solar) \dotfill 
       & 0.6 (0.0--1.5) & 0.5 (0.3--1.1) & 0.6 (0.0--1.3) & 1 (fixed) \\
   $Z_{\rm{Fe}}$ & (solar) \dotfill 
       & 0.2 (0.1--0.8) & 0.2 (0.1--0.6) & 0.2 (0.1--0.4) & 0.5 (0.3--0.7) \\
   $Z_{\rm{Ni}}$ & (solar) \dotfill 
       & 0.2 (0.1--0.8) & 0.2 (0.1--0.6) & 0.2 (0.1--0.4) & 0.5 (0.3--0.7) \\
   $F_{\rm{X}}$\footnotemark[$\ddagger$] & 10$^{-12}$ (erg~s$^{-1}$~cm$^{-2}$) \dotfill 
       & 1.1 (0.5--1.7) & 1.1 (0.7--1.5) & 1.1 (0.7--1.6) & 1.1 (0.8--1.5) \\
   $L_{\rm{X}}$\footnotemark[$\ddagger\S$] & 10$^{34}$ (erg~s$^{-1}$) \dotfill
       & 5.9 & 5.7 & 5.9 & 7.2 \\
   $\chi^{2}$/d.o.f. & \dotfill 
       & 385.2/456 & 391.0/449 & 385.3/447 & 403.3/449 \\
   \hline
   \multicolumn{3}{@{}l@{}}{\hbox to 0pt{\parbox{170mm}{\footnotesize
   \par\noindent
   \footnotemark[$*$] The statistical uncertainties are indicated by the 90\% confidence
   range.
   \par\noindent
   \footnotemark[$\dagger$] Fitting with constraints with $Z_{\rm{Fe}}=Z_{\rm{Ni}}$ (Model 1),
   $Z_{\rm{Fe}}=Z_{\rm{Ni}}$ and $Z_{\rm{Ne}}=Z_{\rm{Mg}}=Z_{\rm{Si}}=Z_{\rm{S}}$ (Model
   2), $Z_{\rm{Fe}}=Z_{\rm{Ni}}$ and $Z_{\rm{N}}=1$ (Model 3), and $Z_{\rm{Fe}}=Z_{\rm{Ni}}$ and $Z_{\rm{N}}=Z_{\rm{O}}=Z_{\rm{S}}=1$ (Model 4).
   \par\noindent
   \footnotemark[$\ddagger$] The X-ray flux in the 0.3--3.0~keV band.
   \par\noindent
   \footnotemark[$\S$] A distance of 13~kpc is assumed.
   }\hss}}
  \end{tabular}
\end{center}
\end{table*}

\medskip

In a transient plasma like that characterizing our V458 Vul spectra, electrons and ions
may not have reached a collisional ionization equilibrium. For non-equilibrium
ionization (NEI) plasmas, low levels of ionization of ions can be seen for the
temperature determined from the electron bremsstrahlung emission (e.g.,
\cite{mewe98}). As a result, the energies of dominant emission features appear shifted
toward lower energies in X-ray CCD spectra by $\gtrsim$10~eV, which is detectable by XIS
\citep{miyata07,bamba08}. The degree of the shifts is parameterized by the ionization
parameter $\int_{0}^{t} dt^{\prime} n_{\rm{e}}(t^{\prime})$, where $n_{\rm{e}}(t)$ is
the electron density as a function of time $t$. We fitted the XIS spectra using an NEI
model \citep{hamilton83,borkowski94,liedahl95,borkowski01} and found a 90\% lower limit
to the ionization parameter of 1.4$\times$10$^{12}$~s~cm$^{-3}$. Finding a value higher
than $\approx 10^{11}$--$10^{12}$~s~cm$^{-3}$ implies that the XIS spectrum is
consistent with being at a collisional equilibrium \citep{mewe98}.

\section{Discussion}\label{s5}
\subsection{Interstellar and Circumstellar Extinction}
From our spectral fitting, we derived a neutral hydrogen column density of $N_{\rm
H}=$~(1.8--4.6)~$\times 10^{21}$~cm$^{-2}$ along the line of sight to V458 Vul
(table~\ref{t2}). The value is consistent with the interstellar extinction estimated by
\citet{poggiani08}. By the time of the Suzaku observation, the color index
(\textit{B}--\textit{V}) of the nova has plateaued from the initial redder
values. Therefore, most of the observed extinction stems from the interstellar medium,
not from the material local to the nova.

The average interstellar hydrogen density ($n_{\rm{H,ISM}}$) in the line of sight is
then estimated as $\sim$0.1~cm$^{-3}$, which is smaller by an order of magnitude than
the canonical value of $n_{\rm{H,ISM}}=$~1~cm$^{-3}$. We suspect that this is due to the
off-plane location of the nova at --3.6~degree in the Galactic latitude
\citep{schlegel98}.

\subsection{Chemical Abundance}
\subsubsection{Previous Abundance Studies}
WDs causing novae can have two different chemical compositions depending on their mass;
CO-type for sources less than $\sim$1.2~$M_{\odot}$ and ONe-type for those more than
1.1--1.2~$M_{\odot}$ \citep{iben89}. With a less mass, the gravitational potential on
the WD surface is smaller and the peak temperature in the thermo-nuclear runaway is
lower. Together with a different core composition, the two types of novae yield
different species of elements in runaways. The major nuclear reaction in CO novae does
not go beyond elements heavier than O, while that in ONe novae reaches up to Si and
S. Theoretical calculations show that intermediate-mass species are enhanced in the
ejecta from ONe novae \citep{starrfield08,jose98,jose07}. Also, the ratio of C and N
against O increases as the WD mass increases \citep{jose98,jose07}. The chemical pattern
of nova ejecta thus provides important clues to guess the type of the WD that hosts the
nova.

Three novae have been studied to discuss the chemical abundance in their hard X-ray
plasma. V382 Vel was observed by medium-resolution gas scintillation proportional
counters aboard BeppoSAX on day 15 \citep{orio01} and by a high-resolution grating
spectrometer aboard Chandra on day 268, after the nova had turned off
\citep{ness05}. V4633 Sgr was observed by medium-resolution CCD spectrometers aboard
XMM-Newton 2.6 and 3.5 years after the explosion \citep{hernanz07}. RS Oph was observed
by high-resolution grating on Chandra and XMM-Newton \citep{nelson08,drake08a,ness08a}
ten times from day~14 to day~239.

In the high-resolution grating spectra of the early hard X-rays in RS Oph
\citep{ness08a}, bremsstrahlung continuum emission and line emission are resolved. N is
found overabundant as a fingerprint of the CNO cycle for the H fusion. High N abundances
thus point to the ejected material from the WD as the source for the X-ray
emission. However, in the specific case of RS Oph, the abundances must be compared to
those of the accreted material from its giant companion (thus the composition of the
companion) in order to confirm that the observed X-rays originate from the WD ejecta. In
a grating spectrum of V382 Vel, a high N abundance was also found
\citep{ness05}. However, the observation was conducted well after the nova had turned
off, thus reflecting the late nebular emission rather than the early hard emission.

In the medium-resolution studies, the chemical pattern is murkier for being unable to
derive the abundances element by element. Nevertheless, \citet{orio01} found a hint of
significant iron depletion in V382 Vel. \citet{hernanz07} tried two thermal models with
different chemical patterns to fit the V4633 Sgr spectra, one with all elements fixed to
the solar values and the other fixed to those of a CO nova shell. Both models can
explain the data.

\subsubsection{Abundance of V458 Vul}
We observed V458 Vul with the medium-resolution spectrometer aboard Suzaku. A high
signal-to-noise ratio spectrum allows us to constrain the abundance of metals
individually for seven elements (table~\ref{t2}). The abundance values are subject to
change due to the unconstrained He abundance. He, along with H, contributes to the
continuum emission in the observed band. We fixed the He abundance to the solar value in
our fitting procedures, but an increased abundance of He leads to an increased estimate
of metal abundances. Therefore, the best-fit abundance values of metals should be
considered only in a sense relative to each other.

Using the result of the ``Model 2'', we define the metal abundance,
$Z_{\rm{m}}=$~0.5~$Z_{\odot}$, which we derived by collectively thawing Ne, Mg, Si, and
S abundances. With respect to this value, the best-fit parameters of the spectrum shows
an enhanced N abundance with $Z_{\rm{N}}$/$Z_{\rm{m}} \sim$~2.6 and deficient O and Fe
abundances with $Z_{\rm{O}}$/$Z_{\rm{m}} \sim$~0 and $Z_{\rm{Fe}}$/$Z_{\rm{m}}
\sim$~0.4.

We note here that systematic uncertainties can arise from possible inaccurate
modeling. First, the uncertainty in the $N_{\rm{H}}$ estimate amplifies the uncertainty
of the N and O abundances, although we consider it unlikely that the both elements have
the solar abundance from the result of the ``Model 4''. Second, a single temperature
model may be an oversimplification. This leads to particularly high uncertainties in
elements formed far away from the isothermal temperature. For example, in a 0.65~keV
plasma, the lines of S and Si (the peak formation temperature $\sim$1.4~keV) have to be
formed in the wings of their line emissivity functions. Small changes in temperature
require large changes in abundance, because the slope of the line emissivity functions
is steeper in their wings than near the peak. This explains the increasing uncertainties
of Mg, Si, to S in the ``Model 1''. The same arguments apply for the low-temperature
lines of N and O (the peak formation temperature $\sim 0.3$~keV). If the plasma is not
isothermal and additional components with temperatures of 0.3 and 1.4 keV are present,
significantly lower abundances for N, O, Si, and S will be obtained. Without a knowledge
of the true emission measure distribution such as that determined by \citet{ness05}, it
is impossible to determine the accurate abundances. We speculate, however, that
contributions from plasma with temperatures below 0.3~keV and above 1.4~keV are minor,
judging from a high ratio of N\emissiontype{VII} to N\emissiontype{VI} and a low ratio
of Si\emissiontype{XIV} and Si\emissiontype{XIII} (figure~\ref{f3}).

With these uncertainties in mind, it is interesting to note that an N enhancement and a
Fe deficiency in hard X-ray emission are seen in all novae with a $Z_{\rm{N}}$ and a
$Z_{\rm{Fe}}$ measurement, respectively, including the present work
\citep{mukai01,orio01,ness05,ness08a}. We speculate that this typical pattern in the
X-ray spectra is a consequence of these novae belonging to the ONe-type. The deficiency
of the Fe abundance with respect to the intermediate-mass elements is consistent with
the theoretical understanding that ONe-type novae produce elements up to Si and S. The N
enhancement can be interpreted as an outcome of the H fusion on the WD surface, where
the CNO cycle produces abundant N by the N to O conversion as the rate-determining
process The abundance of N with respect to O increases for more massive ONe-type
novae. A quantitative comparison with theoretical work
\citep{starrfield08,jose98,jose07} is hampered by a relatively large uncertainty in our
fitting results, but the overall abundance pattern in V458 Vul indicates that this is an
ONe-type nova.

\subsection{Plasma Origin and Parameter Constraints}
We first argue that the X-ray emission observed from V458 Vul is most likely to be the
internal shock of the nova ejecta, and not due to other processes. The hard spectrum
exceeding 1~keV shows that it is neither the SSS emission nor the reprocessed emission
of it. The possible N enhancement is consistent with the plasma mostly composed of the
ejecta. Shocks might occur externally between the ejecta and swept-up interstellar
matter, but the total mass of the swept-up matter by the time of the Suzaku observation
is too small compared to the ejected mass. The ejecta mass is estimated to be in the
range of $\sim$10$^{-3}$--10$^{-6}$ $M_{\odot}$ \citep{gehrz98}, whereas the swept-up
mass is $\sim 4\pi/3 (v_{0}t_{\rm{S}})^3 m_{\rm{p}} n_{\rm{H,ISM}} \sim 10^{-11}
M_{\odot}$ where $m_{\rm{p}}$ is the proton mass, $n_{\rm{H,ISM}} = 1$~cm$^{-3}$ is the
ISM density, $v_{0} = 1.8 \times 10^{3}$~km~s$^{-1}$ is the initial speed of expanding
shell \citep{buil07}, and $t_{\rm{S}}$ is the time of the Suzaku observation elapsed
from the explosion. If we adopt $n_{\rm{H,ISM}} = 155$~cm$^{-3}$, the densest value in
the surrounding planetary nebula \citep{wesson08}, the swept-up mass is far smaller than
the typical ejecta mass.

In the case of symbiotic novae like RS Oph, early hard emission can be produced by
external shocks between the expanding ejecta and the stellar wind of the giant companion
\citep{bode06}. This does not work for classical novae, in which the secondary is a main
sequence star and the stellar wind is not dense enough \citep{obrien94}. While no
observational evidence exists that V458 Vul is not a symbiotic nova, shock emission by
collisions between the expanding ejecta and the stellar wind of the companion is not
likely. In RS Oph, the shock itself occurred over a relatively short time scales, and
the hard X-ray emission quickly faded afterwards \citep{bode06}, while the X-ray count
rate in V458 Vul rises with no indication of a decline. It is most likely that the
X-rays from V458 Vul originate in shocks internal to the expanding system, presumably
between slower and earlier pre-maximum ejecta and faster and later nova winds
\citep{friedjung87,mukai01}. However, in the later Swift observations one year after the
nova, this hard X-ray emission still remains and exhibits highly variable flux in
anti-correlation with the UV flux \citep{ness08b}. Elaborated theoretical work is
necessary to explain all these properties to see if the internal shock interpretation is
valid.

From the spectral fits, the absorption-corrected X-ray luminosity at the Suzaku
observation is $L_{\rm{X}} (t_{\rm{S}})= 6 \times 10^{34}$~erg~s$^{-1}$ in the
0.3--3.0~keV band at an assumed distance of 13~kpc. The luminosity is in a typical
range for classical novae \citep{mukai08}. The X-ray volume emission measure at the same
time is $EM(t_{\rm S}) = n_{\rm{e}}^{2}(t_{\rm{S}}) V_{\rm{X}}(t_{\rm{S}}) \sim 7
\times 10^{57}$~cm$^{-3}$, where $V_{\rm{X}}(t)$ and $n_{\rm{e}}(t)$ represent the
X-ray-emitting volume and the plasma density as a function of time $t$.

With these measurements and several simplifying assumptions, a crude estimate can be
obtained for the lower and upper bounds of the plasma density at $t_{\rm{S}}$. Assuming
that the plasma density is uniform and that the expansion velocity $v_{0}$ has not
changed since the outburst, the X-ray emitting volume cannot exceed the volume of the
expanding shell; i.e., $V_{X}(t_{\rm{S}}) < 4\pi/3 (v_0 t_{\rm{S}})^{3}$. The observed
$EM(t_{\rm{S}})$ thus implies that the density must be $n_{\rm{e}}(t_{\rm{S}}) > 8
\times 10^{5}$~cm$^{-3}$. If we further assume that the plasma was created at one epoch
and not heated repeatedly, the plasma density cannot be too high to cool radiatively too
soon. The cooling time $t_{\rm rc}$ at $t_{\rm{S}}$ is approximated as
$3n_{\rm{e}}(t_{\rm{S}})k_{\rm{B}}T(t_{\rm{S}})V_{X}(t_{\rm{S}})/L_{\rm{X}}(t_{\rm{S}})$.
With $t_{\rm rc} >$ 100~day, which is an approximate duration of the detected X-ray
coverage by Swift, the upper bound is estimated as $n_{\rm{e}}(t_{\rm{S}}) < 4 \times
10^{7}$~cm$^{-3}$. The lower and upper bounds put $n_{e}(t_{\rm{S}})$ within a range of
two orders around $\approx$10$^{6}$~cm$^{-3}$. \citet{orio96} derived a similar density
of $n_{\rm{e}} \sim 2 \times 10^{6}$~cm$^{-3}$ assuming $V_{\rm{X}}(t) = 4\pi/3 (v_0
t)^{3}$ for V351 Pup observed 16 month after its explosion.

\section{Summary}\label{s6}
The classical nova V458 Vul was discovered on 2007 August 8. In response to the Swift
detection of X-ray emission on day 70, we conducted a target of opportunity observation
on day 88 with Suzaku.

The X-ray spectrum obtained by the XIS is attributable to a single-temperature
($\sim$0.64~keV) optically-thin thermal plasma emission at a collisional equilibrium
with an interstellar extinction of $\sim$3$\times$10$^{21}$~cm$^{-2}$. In addition to
the plasma temperature and the amount of extinction, the chemical abundances of
conspicuous elements were derived. The abundance pattern is characterized by a possible
enhancement of N and deficiencies of O and Fe abundances with respect to other
metals. From the X-ray luminosity of $6 \times 10^{34}$~erg~s$^{-1}$, the plasma
temperature of 0.64~keV, and the volume emission measure of $7 \times 10^{57}$~cm$^{-3}$
at the time of the Suzaku observation, we constrained the average plasma density to be
$8 \times 10^{5}$--$4 \times 10^{7}$~cm$^{-3}$ assuming a constant expansion velocity of
$1.8\times 10^{3}$~km~s$^{-1}$. All the observed X-ray properties obtained by Suzaku can
be interpreted as the emission arising from shocks of ejecta in a nova at a $\sim$13~kpc
distance.

We demonstrated the capability of the XIS to yield a high signal-to-noise ratio spectrum
for a moderately bright X-ray nova in a reasonable telescope time. This put constraints
on physical values and chemical compositions of the X-ray emission. In conjunction with
monitoring observations routinely done by Swift, we anticipate that similar Suzaku
observations can be executed for several novae every year.

\bigskip

The authors appreciate an excellent review by Gloria Sala. We also thank Kazuhiro
Nakajima and Hitoshi Yamaoka in the Variable Star Observers League in Japan for
providing optical data of V458 Vul, and the Suzaku telescope managers for allocating a
part of the director's discretionary time for this observation. Support for this work
was provided by the National Aeronautics and Space Administration through Chandra
Postdoctoral Fellowship Awards Number PF6-70044 (M.\,T.) and PF5-60039 (J.-U.\,N.) 
issued by the Chandra X-ray Observatory Center, which is operated by the Smithsonian
Astrophysical Observatory for and on behalf of the National Aeronautics Space
Administration under contract NAS8-03060. J.\,J.\,D. was supported by NASA contract
NAS8-39073 to the CXC during the course of this research. D.\,T. is financially
supported by the Japan Society for the Promotion of Science. This research has made use
of the SIMBAD database operated at CDS, Strasbourg, France and the NASA/IPAC
Extragalactic Database (NED) which is operated by the Jet Propulsion Laboratory,
California Institute of Technology, under contract with the National Aeronautics and
Space Administration.


\end{document}